# $Pt$, $Ni$ and $Ti$ Schottky barrier contacts to $\beta$-$(Al_{0.19}Ga_{0.81})_2O_3$ grown by Molecular Beam Epitaxy on Sn doped β-$Ga_2O_3$ substrate

Abhishek Vaidya[1,a)], K. Sasaki[2], A. Kuramata[2], T. Masui[2], and Uttam Singisetti [1b)]

[1]*Electrical Engineering Department, University at Buffalo, Buffalo, NY, 14260, USA*

[2] *Novel Crystal Technology, Inc., Sayama, Saitama 350-1328, Japan*

A comprehensive current-voltage $(I-V)$ characterization is performed for three different Schottky contacts; $Pt$, $Ni$ and $Ti$, to unintentionally doped (UID) $\beta$-$(Al_{0.19}Ga_{0.81})_2O_3$ grown by molecular beam epitaxy (MBE) on $\beta$-$Ga_2O_3$ for temperatures ranging between 25 ℃ – 300 ℃. Reciprocal space mapping shows the $(Al_{0.19}Ga_{0.81})_2O_3$ films are strained and lattice matched to the substrate. Schottky Barrier Height ($SBH$), ideality factor ($n$), and series resistance ($R_s$) are extracted from the $I-V$ characteristics for the three types of metals and temperatures. Room temperature capacitance-voltage $(C-V)$ measurements revealed fully depleted $\beta$-$(Al_{0.19}Ga_{0.81})_2O_3$ layer. Extracted room temperature $SBH$s after zero field correction for $Pt$, $Ni$ and $Ti$ were 2.39 $eV$, 2.21 $eV$, and 1.22 $eV$ respectively. Variation of $SBH$s with metal clearly indicates the dependence on work function.

---

a) Email: avaidya4@buffalo.edu, Tel: 716-645-1017

b) Email: uttamsin@buffalo.edu, Tel: 716-645-1536

$\beta$-Gallium oxide ($\beta$-$Ga_2O_3$) has become a material of prime interest for power electronic device research in recent years due to its attractive properties such as ultra-wide band gap ($E_g \sim 4.8 - 4.9\ eV$),[1,2] very high Baliga's Figure of Merit (BFoM)[3,4], and a high breakdown field strength of $8\ MV/cm$.[5,6] $\beta$-$Ga_2O_3$ ($Ga_2O_3$) substrates are grown by scalable bulk crystal growth technology which could lead to lower device costs.[3,7,8] Several groups have reported $Ga_2O_3$ based field effect transistors (FETs) in recent years which shows the potential of the $Ga_2O_3$ technology.[6,9-11] Although it has high BFoM, the electron mobility is low [12,13] when compared to the standard power semiconductors such as $Si$, $GaN$, $SiC$. One approach to increase the mobility is to use a two-dimensional electron gas (2-DEG) in an $(Al_xGa_{1-x})_2O_3/Ga_2O_3$ heterostructure similar to the *AlGaAs/GaAs* heterostructure. The idea behind such structures is to physically separate the carriers from the dopants in order to reduce the impurity scattering[14], recent calculations have shown significant improvement in the mobility.[15] The barrier layer in these heterostructures is $\beta$-$(Al_xGa_{1-x})_2O_3$ which is lattice matched to the $\beta$-$Ga_2O_3$. Several groups have recently reported the existence of the 2-DEG in $(Al_xGa_{1-x})_2O_3/Ga_2O_3$ heterostructures and the transistor characteristics.[16-19] The Schottky contacts to this $(Al_xGa_{1-x})_2O_3$ barrier serves as the gate contact in these devices. Metal-semiconductor contacts are important components of such high electron mobility devices; thus it becomes important to evaluate different parameters related to Schottky contacts. Several groups have reported Schottky contacts to $Ga_2O_3$[20-23] and Ahmadi *et al.*[24] have reported $Ni$ Schottky contacts on $(Al_xGa_{1-x})_2O_3$ for varying *x* values up to 16%. However, a detailed contact $I - V$ study with different metals has not been reported till date. A detailed Schottky contact $I - V$ study is instrumental in making the selection of metals for optimum device current-voltage ($I - V$) characteristics in a transistor. Schottky barrier contacts are characterized primarily by four parameters: Schottky barrier heights (*SBH*s), ideality factors (*n*), series resistances ($R_s$) and reverse saturation current ($I_s$).

In this letter, we report the room temperature and high temperature characteristics of *Pt*, *Ni* and *Ti* contacts to unintentionally doped (UID) *β-(Al<sub>0.19</sub>Ga<sub>0.81</sub>)<sub>2</sub>O<sub>3</sub>*. *SBH*, *n*, $R_s$ for each contact are calculated by using the method proposed by Cheung *et al.*[25], which is widely used for Schottky contact characterization. However, the extracted *SBH* ($\Phi_{B,nc}$) obtained using this popular method is affected by the presence of the electric field in the structure, and also by the non-ideal effects in the diode. These effects need to be corrected to obtain the true zero electric field barrier height ($\Phi_{B,zf}$) as proposed by Wagner *et al.*[26] It is necessary to use the later method in this study because of the high ideality factor (*n*) observed for all the Schottky contacts. In addition the *SBH* ($\Phi_{B,vbi}$) was also extracted from built in voltage and was compared with $\Phi_{B,zf}$.

For the diode fabrication, nominally 115 *nm* UID - *(Al<sub>0.19</sub>Ga<sub>0.81</sub>)<sub>2</sub>O<sub>3</sub>* was grown by molecular beam epitaxy (MBE) on *Sn* doped *β-Ga<sub>2</sub>O<sub>3</sub>* substrates (010). The (UID) *β-(Al<sub>0.19</sub>Ga<sub>0.81</sub>)<sub>2</sub>O<sub>3</sub>* layer was grown at 555 °C, with Ga cell at 770 °C and the Al cell at 850 °C and a mixture of $O_2$ and $O_3$ gases under 2 sccm flow rate.[18] Fig. 1(a) shows the reciprocal space map of a structure grown under same conditions, which shows the films are lattice matched to the substrate. Fig. 1(b) shows the atomic force microscopy (AFM) scan of the epitaxial layer which has root mean square (RMS) roughness of 0.537 *nm*. After standard solvent cleaning procedure, the back side of sample was treated with a short BCl<sub>3</sub>/Ar reactive ion etching process for enhanced ohmic contact formation before blanket depositing *Ti* (20 *nm*) / *Au* (70 *nm*) as the cathode metal. On the (UID) *β-(Al<sub>0.19</sub>Ga<sub>0.81</sub>)<sub>2</sub>O<sub>3</sub>* side of the sample, circular contacts with a 50 μ*m* radius were defined by standard electron beam lithography procedure and *Ti* (20 *nm*)/ *Au* (70 *nm*) lift-off process. The process was repeated for *Ni* and *Pt*. 70 *nm Au* ensures good probing and serves as a protective layer for the underlying Schottky metal. A schematic of the final structure and a representative energy band diagram of the metal-(UID) β-(Al$_{0.19}$Ga$_{0.81}$)$_2$O$_3$ is shown in Fig. 2(a-b). Finally, all the contacts were annealed simultaneously in a rapid thermal anneal (RTA) chamber at 400 °C for 1 minute in a Nitrogen ambient to improve the ohmic contact. It is noted that annealing reduced the ohmic contact resistance and the series resistance in the Schottky diodes.

Next, $I-V$ measurements were carried out using HP 4155B semiconductor parameter analyzer. A Temptronics thermal chuck was used for temperature control of the probing stage. Bias voltage was swept from -10 $V$ to 4 $V$ with a current compliance limit of 40 $mA$ and a step size of 20 $mV$ was used in order to get sufficient data points in the forward bias region. The measurements were repeated for 100 °C, 200 °C, 250 °C and 300 °C. A standard reverse biased $C-V$ measurement was also performed on the Schottky contacts using Agilent 4294A precision impedance analyzer for extracting the doping density and extraction of $SBH$. However, the $C-V$ measurement showed capacitance values of the order of few Pico farads and a modulation of few fF/volt with increasing reverse bias (see Supplementary Material), which indicates a completely depleted (UID) β-(Al$_{0.19}$Ga$_{0.81}$)$_2$O$_3$ layer. Similar results have been reported previously in ref 24. Consequently, $C-V$ method for $SBH$ extraction was not used in further analysis. A room temperature reverse breakdown measurement was also performed which presented a very low breakdown voltage of ~ 75 $V$, further tests are necessary to understand the low breakdown voltage.

Room temperature current density ($J$) – voltage ($V$) curves for the $Pt$, $Ni$ and $Ti$ Schottky contacts are plotted in Fig. 3(a). It is clear from the $J-V$ plots that all the three metals form Schottky contacts with the (UID) β-(Al$_{0.19}$Ga$_{0.81}$)$_2$O$_3$ layer, with clearly distinct turn on voltages, indicating dependence of barrier height on the choice of metal. An ON-OFF current ratio as high as $10^9$ has been achieved for $Pt$ (See Fig. 4(a) ) and $Ni$ (See Fig. 4(c) ) contacts at room temperature. Forward current densities of 100 $A/cm^2$ have been achieved at bias voltages of 2.45 V, 2.23 V and 1.23 V with specific resistance values of 1.8 $m\Omega \cdot cm^2$, 1.6 $m\Omega \cdot cm^2$ and 1.9 $m\Omega \cdot cm^2$ for $Pt$, $Ni$ and $Ti$ respectively (see Table I). Low reverse bias current values of $Pt$ and $Ni$ also show that they have excellent rectifying behavior; the reverse current densities of $Pt$ and $Ni$ are below $10^{-7}$ $A/cm^{-2}$. However the $Ti$ contacts have approximately five orders of magnitude larger reverse current densities (See Fig. 4(e) ). This behavior is typically seen for low barrier $Ti$ Schottky contacts.

For the Schottky contacts, current-voltage behavior is represented mathematically by the thermionic emission current equation (eq. 1)

$$I = I_s \left\{ \exp\left[\frac{q(V - IR_s)}{nk_BT}\right] - 1 \right\} \quad (1)$$

where $q$ is the electronic charge, $n$ is the ideality factor, $k_B$ is the Boltzmann constant, $R_s$ is the series resistance and $I_s$ is the reverse saturation current density given by

$$I_s = A_{eff} A^{**} T^2 \exp\left(-\frac{q\Phi_{B,nc}}{kT}\right) \quad (2)$$

where $A_{eff}$ is the area of the contact, $A^{**}$ is the Richardson constant and $\Phi_{B,nc}$ is the barrier height before zero field correction. For (UID) $\beta$-$(Al_{0.19}Ga_{0.81})_2O_3$, we use electron effective mass of 0.342 with a corresponding $A^{**} = 41.1 \frac{A}{cm^2 \cdot K^2}$.[24] For $V \gg kT$ we can ignore the $-1$ in diode current equation and it can be rewritten in the form of eq. 3

$$V = R_s A_{eff} J + n\Phi_{B0} + \frac{n}{\beta} \ln \frac{J}{A^{**}T^2} \quad (3)$$

where $\beta = q/kT$. Above equation is differentiated with respect to $J$ and rearranged as eq. 4, a linear curve fit to which gives both $R_s$ & $n$ from the slope and y-intercept respectively as indicated in Fig. 3(b).

$$\frac{dV}{d(\ln J)} = R_s A_{eff} J + \frac{n}{\beta} \tag{4}$$

Room temperature $n$ and $R_s$ for $Pt$, $Ni$ and $Ti$ obtained using this method are listed in Table I. Ideality factor ($n$) quantifies the current transport mechanisms in a metal-semiconductor junction[27], with $n = 1$ representing a pure thermionic emission current and an ideal interface with homogeneous barrier height; while $n > 1$ represents thermionic field emission and/or trap-assisted tunneling currents, non-homogeneous barrier height could be involved along with thermionic emission. Thermionic field emission is the tunneling of thermally excited carriers which see a thinner barrier compared to tunneling.[28] Apart from this, metal adhesion quality and interface quality also affect the ideality factor. The larger the value of $n$, more dominant are the non-ideal effects. All the Schottky contacts in this experiment have $n$ values far greater than 1 as seen in Table I, which could be treated as a combined effect of all the non-ideal phenomena. However, Schottky contacts to $Ga_2O_3$ have shown ideality factor of ~1.[20]

To find the $SBH$ ($\Phi_{B,nc}$), the function $H(J)$ in eq. 5 can be defined by rearranging terms in eq. 3. Using the linear curve fit to $H(J)$ vs $J$ plot, y-intercept gives the value of $\Phi_{B,nc}$ as shown in Fig 3(b).

$$H(J) = V - \frac{n}{\beta} \ln \frac{J}{A^{**}T^2} = R_s A_{eff} J + n\Phi_{B,nc} \tag{5}$$

The accuracy of $SBH$ obtained from this method depends on the corresponding value of ideality factor and is closer to the actual $SBH$ if $n$ is close to unity as described by Wagner et al.[26] However, if $n$ is large, which is the case in this study, a correction is necessary to obtain the true zero field SBH ($\Phi_{B,zf}$). It is obtained by using the following equation [26]

$$\Phi_{B,zf} = \left[\Phi_{B,nc} - \left(\frac{n-1}{n}\right)\frac{kT}{q}ln\frac{N_C}{N_D}\right]n \qquad (6)$$

where $N_D$ is taken to be the donor concentration in β-Ga$_2$O$_3$ since C-V data indicated completely depleted UID and the fermi level is essentially controlled by the doping levels in the substrate. $N_C$ is the conduction density of states for β-Ga$_2$O$_3$ calculated using electron effective mass of 0.342 with all other constants having their standard values. The extracted room temperature $\Phi_{B,zf}$ values for the $Pt$, $Ni$ and $Ti$ contacts 2.39 $eV$, 2.21 $eV$ and 1.22 $eV$ respectively. The SBH values are larger than that of Ga$_2$O$_3$ which is expected due to the larger bandgap of *β-(Al$_{0.19}$Ga$_{0.81}$)$_2$O$_3$*.[20,21] Table I lists the calculated *SBH* using different methods. Ideal *SBH* is governed by metal work function and electron affinity of semiconductor. Assuming the electron affinity of the semiconductor to be constant throughout the study, barrier height should increase with increasing metal work function which agrees with the presented $\Phi_{B,zf}$ values in Table I. It is noted that room temperature *SBH* value reported for $Ni$ Schottky contact on *(Al$_{0.164}$Ga$_{0.836}$)$_2$O$_3$* (Ahmadi et al.[24]) is much lower compared to $\Phi_{B,zf}$ value in this study and is comparable to $\Phi_{B,nc}$. Considering both the reported contacts have high ideality factors, the need for using zero field correction method (eq. 6) to characterize the intrinsic metal-semiconductor interface is elucidated. To further verify these values, another method which involves determining *SBH* from built-in voltage ($V_{bi}$) and doping concentration ($N_D$) is used in this study. $V_{bi}$ was extracted from linear forward $J-V$ plot by linear curve fit as show in Fig. 3(a), the $V_{bi}$ value for which is the $x$ intercept and corresponding barrier height ($\Phi_{B,vbi}$) is then calculated using eq. 7 ( ignoring small voltage drop at the *(Al$_{0.19}$Ga$_{0.81}$)$_2$O$_3$*/Ga$_2$O$_3$ interface)

$$\Phi_{B,vbi} = V_{bi} + \frac{kT}{q}\ln\left(\frac{N_C}{N_D}\right) \tag{7}$$

Room temperature $\Phi_{B,vbi}$ for each contact are listed in Table I and it can be observed that they are comparable to room temperature $\Phi_{B,zf}$ values.

Fig. 4(a) - 4(f) show temperature dependent trends of $J$, $n$, $\Phi_{B,zf}$ and $\Phi_{B,vbi}$ for each metal. With temperature increasing from 27 °C to 250 °C forward current densities increase for a constant voltage, for all contacts as expected due to the reduction of the $SBH$. An apparent delayed turn on observed in Fig. 4(a) is due to resolution and measurement accuracy limitations of the I-V measurement setup for the actual $Pt$ Schottky contact. Observed reduction in the values of $n$ with temperatures up to 250 °C could be due to marginal improvement in the pure thermionic emission current part over thermionic field emission current in the device while the net ideality factor still remains much above 1 indicating broadly unaffected factors such as interface quality and SBH inhomogeneity. However, as the temperature increases above 250 °C the series resistance increases leading to increased $I \cdot R_s$ drop near and above $V_{bi}$ values reducing the current densities. Fig 5(a) shows the extracted $R_s$ vs Temperature, which shows a steep increase in series resistance after 250 °C. The increased $I \cdot R_s$ voltage drop from 250 °C to 300 °C can also be noticed in the linear temperature dependent $J-V$ plot of Fig. 5(b). The series resistance is a combination of contact and bulk resistance. Bulk resistance increases with increasing temperature due to the decrease in mobility and the contact resistance decreases with increasing temperature in Ga$_2$O$_3$. The overall trend is determined by two individual trends for these resistances. We attribute that bulk resistance dominates beyond 250 °C and hence we see a sudden increase in the series resistance. Such non-monotonic behavior for parasitic source resistance has been observed in previous studies on β-Ga$_2$O$_3$ transistors.[29] Barrier heights extracted by both methods ($\Phi_{B,zf}$ and $\Phi_{B,vbi}$) reduce with increasing temperature which indicates they are corresponding to thermionic emission since pure thermionic emission barrier will have reduced barrier heights at elevated temperatures.[30] The slight increase of the *SBH* at 300 °C is due to the large series resistance; Fig. 5(b) clearly shows the increase in $V_{bi}$ due to the increased series resistance drop at 300 °C. Reverse current densities increase monotonically with temperature for all the contacts. This is attributed to the increase in electron energy at higher temperatures which enables them to climb over the barrier from metal to semiconductor and also this increased electron energy leads to a lower barrier height for thermionic field emission.[28] As a result ON-OFF current ratios is reduced by two to three orders of magnitude for Pt, Ni and eight orders of magnitude for Ti due to exponential increase in the reverse current densities with temperature.

In summary, we extracted the *SBH* ($\Phi_{B,zf}$), $n$ and $R_s$ from forward $I-V$ characteristics of three different metal-(UID) *β-(Al$_{0.19}$Ga$_{0.81}$)$_2$O$_3$* Schottky contacts for a range of temperatures. $Pt$ and $Ni$ show promising Schottky contact behavior with large *SBH* favorable for high power devices. In general, $n$ and $\Phi_{B,zf}$ show a negative temperature coefficient which is attributed to increased thermionic emission current and increased thermal energies of electrons respectively. The zero field correction approach gives intrinsic SBH for the metal-(UID) *β-(Al$_{0.19}$Ga$_{0.81}$)$_2$O$_3$* contacts. Higher $n$ values in the Schottky contacts is attributed to various non-ideal effects.

See Supplementary Material for the capacitance-voltage profile of all the contacts and the Richardson plot.

## ACKNOWLEDGMENTS

This work is supported by National Science Foundation (NSF) grant (ECCS 1607833) monitored by Dr. Dimitris Pavlidis. A portion of this work was performed in University at Buffalo, Davis Hall Electrical Engineering Cleanroom. The authors would like to thank the support from cleanroom staff.

**FIGURE CAPTIONS:**

Fig. 1. (a) A (110) RSM of the *β-(Al$_x$Ga$_{1-x}$)$_2$O$_3$* grown on β-Ga$_2$O$_3$ substrate, showing that the *(Al$_x$Ga$_{1-x}$)$_2$O$_3$* epilayer was coherently strained to the underlying β-Ga$_2$O$_3$ substrate. A reciprocal lattice unit (rlu) corresponds to λ/2d, where λ is the x-ray wavelength and d is the interplanar spacing. The thickness (116 nm) and composition of the *(Al$_x$Ga$_{1-x}$)$_2$O$_3$* layer was measured using Rutherford back scattering (RBS) (b) AFM scan image of the *(Al$_{0.19}$Ga$_{0.81}$)$_2$O$_3$* surface showing average and RMS surface roughness $0.538\ nm$.

Fig. 2. (a) Schematic Cross-section of Schottky diode fabricated on 115 $nm$ (UID) *β-(Al$_{0.19}$Ga$_{0.81}$)$_2$O$_3$/β-Ga$_2$O$_3$* substrate (b) Representative zero bias energy band diagram of metal-semiconductor junction

Fig. 3. (a) Room temperature current density ($J$) $vs$ voltage ($V$) plots for $Pt$, $Ni$ and $Ti$ Schottky contacts showing distinct turn-on voltages as a function of metal work function. $V_{bi}\ fit$ shows the $V_{bi}$ extraction method. (b) $\frac{dV}{d(\ln J)}$ $vs\ J$ and 'fit 1' curve fitting, Y-intercept (left) gives $n$ and slope gives $R_s$. H $vs\ J$ and 'fit 2', Y-intercept (right) gives $\phi_{B,nc}$ and slope gives $R_s$.

Fig. 4. $\log J\ vs\ V$ plots at different temperatures showing variation in forward and reverse current densities with temperature for (a) *Pt- β-(Al$_{0.19}$Ga$_{0.81}$)$_2$O$_3$* (c) *Ni- β-(Al$_{0.19}$Ga$_{0.81}$)$_2$O$_3$* (e) *Ti- β-(Al$_{0.19}$Ga$_{0.81}$)$_2$O$_3$*. Comparison of $n$ and SBHs extracted using two different methods with temperature for (b) *Pt- β-(Al$_{0.19}$Ga$_{0.81}$)$_2$O$_3$* (d) *Ni- β-(Al$_{0.19}$Ga$_{0.81}$)$_2$O$_3$* (f) *Ti- β-(Al$_{0.19}$Ga$_{0.81}$)$_2$O$_3$*.

Fig. 5. (a) Extracted $R_s$ vs $T$ (°C) plot for $Pt$, $Ni$ and $Ti$ contacts showing steep increase in resistance beyong 250 °C in all the three contacts. (b) Illustrative linear $J$ vs $V$ plot for $Pt$ contact showing $R_s$ effect on forward characteristics.

**FIGURES**

Fig 1:

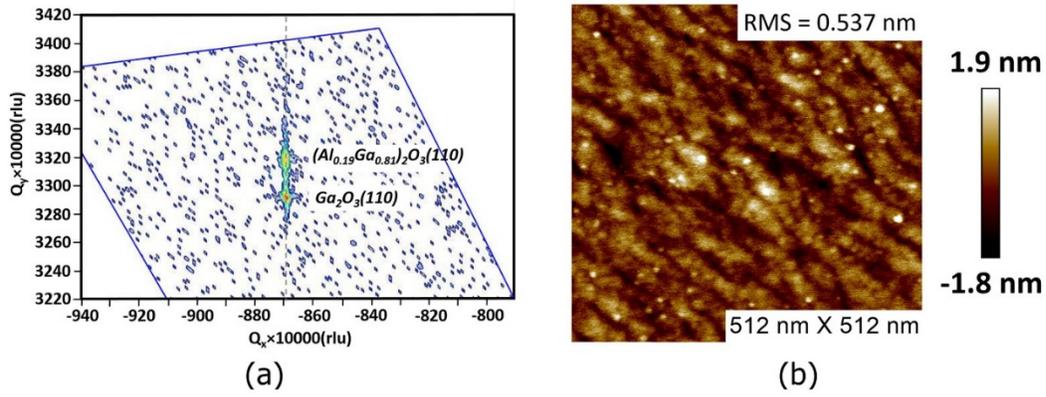

Fig 2:

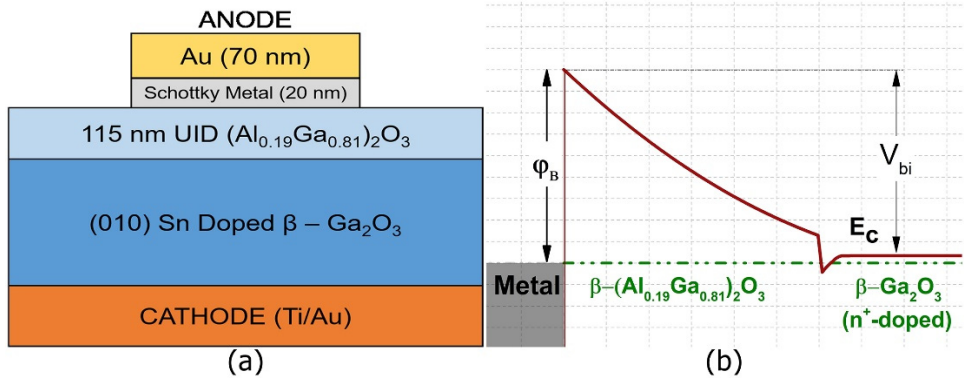

Fig 3:

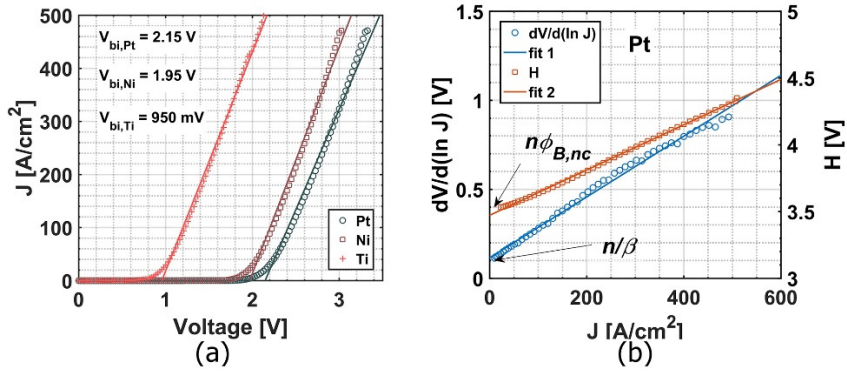

Fig 4:

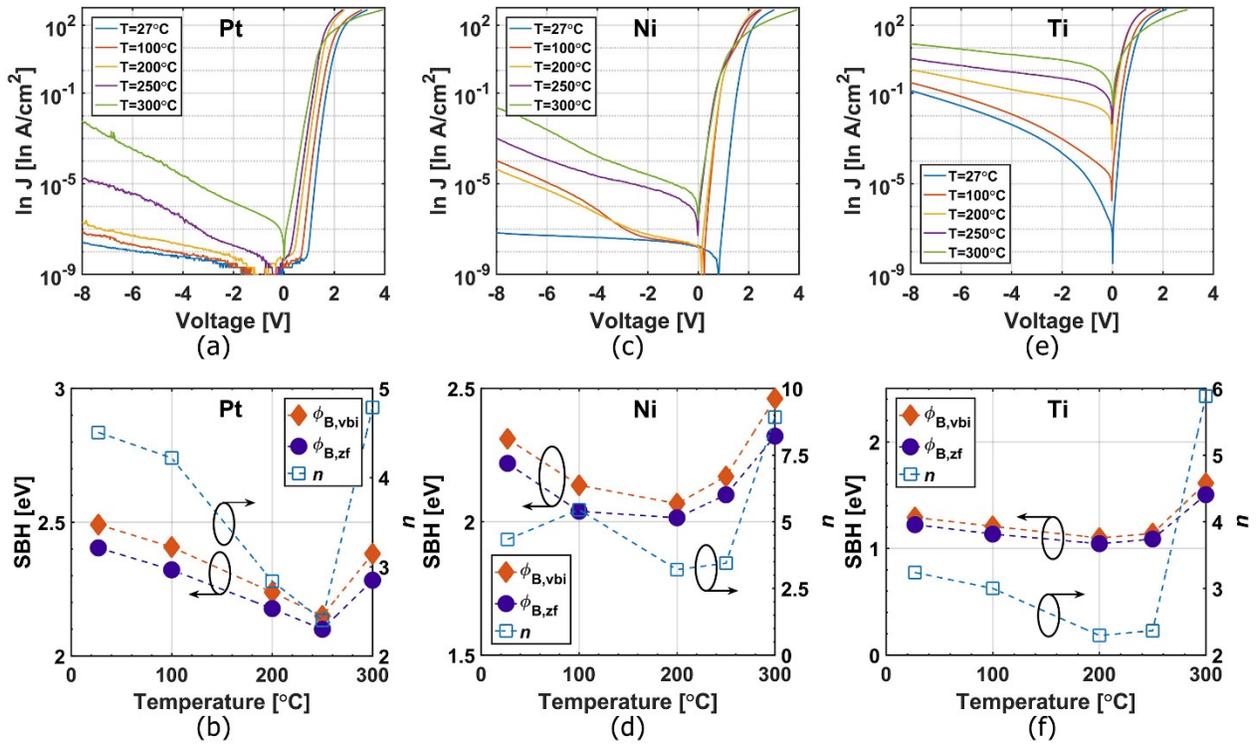

Fig 5:

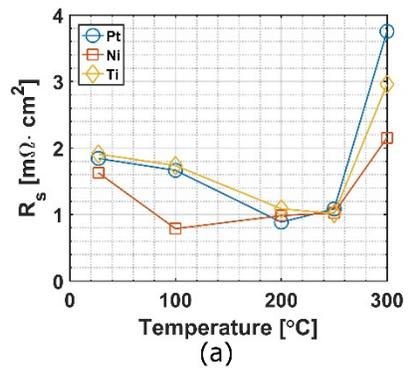
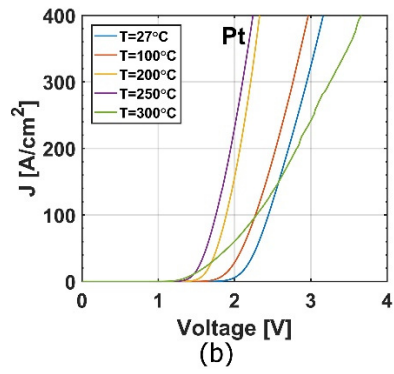

(a) (b)

TABLE CAPTIONS:

Table I. Extracted room temperature values of $SBH$ from buil-in voltage ($\Phi_{B,vbi}$), from zero-field condition ($\Phi_{B,zf}$) and without zerofield correction ($\Phi_{B,nc}$), ideality factor ($n$) and series resistance ($R_s$) for $Pt$, $Ni$ and $Ti$ contacts.

TABLES:

Table I.

| Metal | $n$ | $\Phi_{B,zf}(eV)$ | $\Phi_{B,vbi}(eV)$ | $\Phi_{B,nc}(eV)$ | $R_s(m\Omega \cdot cm^2)$ |
|---|---|---|---|---|---|
| **Platinum** | 4.45 | 2.39 | 2.49 | 0.779 | 1.8 |
| **Nickel** | 4.27 | 2.21 | 2.30 | 0.756 | 1.6 |
| **Titanium** | 3.23 | 1.22 | 1.29 | 0.59 | 1.9 |

# Supplementary Material for

## $Pt$, $Ni$ and $Ti$ Schottky barrier contacts to $\beta$-$(Al_{0.19}Ga_{0.81})_2O_3$ grown by Molecular Beam Epitaxy on Sn doped β-Ga$_2$O$_3$ substrate

Abhishek Vaidya[1,a)], K. Sasaki[2], A. Kuramata[2], T. Masui[2], and Uttam Singisetti [1b)]

[1]*Electrical Engineering Department, University at Buffalo, Buffalo, NY, 14260, USA*

[2] *Novel Crystal Technology, Inc., Sayama, Saitama 350-1328, Japan*

## I. Capacitance-Voltage measurement data for Pt, Ni and Ti Schottky diodes on *UID β-(Al$_{0.19}$Ga$_{0.81}$)$_2$O$_3$.*

Fig. S1(a-f) shows room temperature reverse bias C-V data and corresponding depth vs carrier density profile for all the Schottky diodes associated with this study. The depth profile confirms the fully depleted UID layer and shows carrier density profile in the heavily doped *β-Ga$_2$O$_3$* substrate.

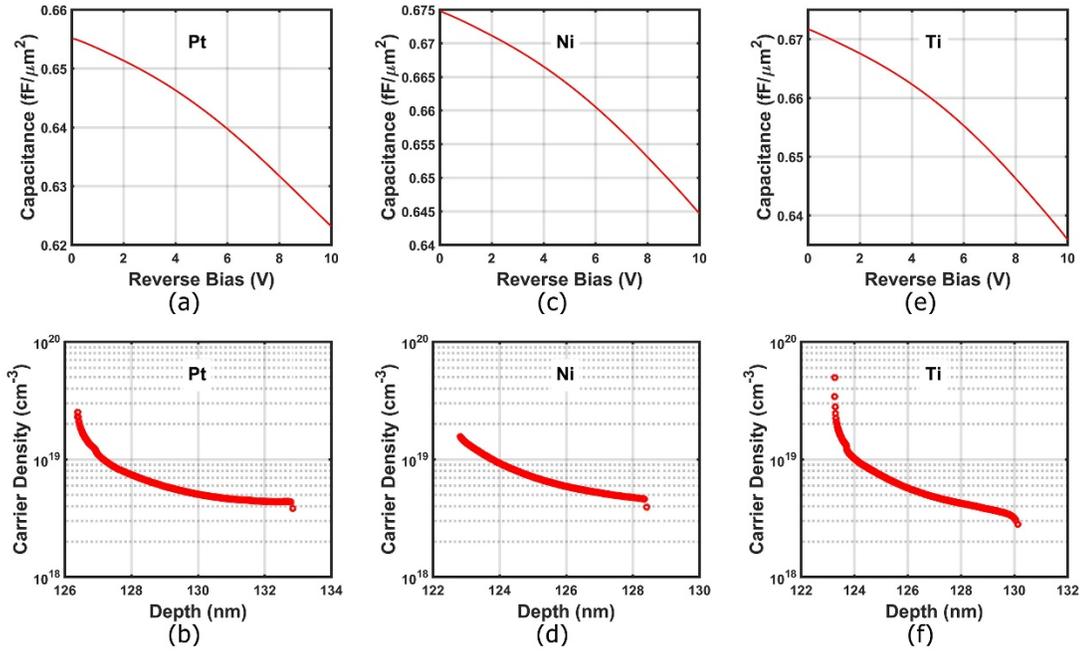

FIG. S1. Room temperature reverse bias C-V data for Pt (a), Ni(c), Ti(e) and corresponding depth vs carrier density profiles (b, d, f respectively).

## II. Richardson plot for the Pt, Ni, and Ti Schottky contacts

Figure below shows the Richardson plot for Pt, Ti and Ni Schottky diodes. The values of SBHs so obtained are significantly lower compared to $\Phi_{B,zf}$ and $\Phi_{B,vbi}$ values presented in the Table I for corresponding metals. This is probably due to temperature dependence of SBH and ideality factor since the Richardson plot method assumes both SBH and ideality factor constant over a small temperature range around room temperature.

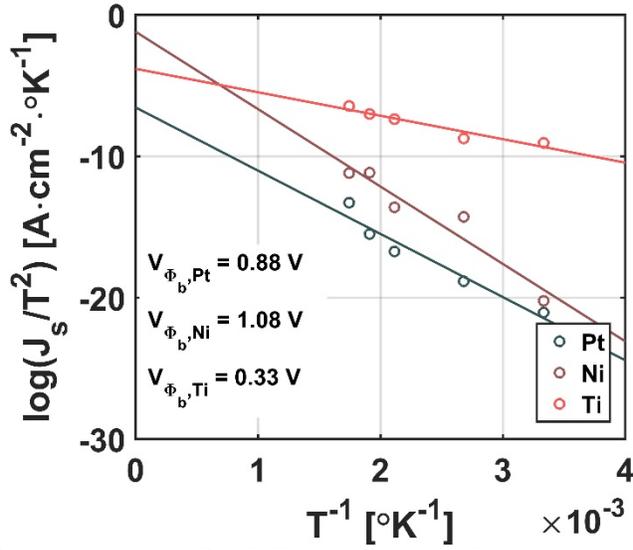

FIG. S2. Richardson plot for Pt, Ni, Ti contacts, where $J_s$ is reverse saturation current density for the Schottky diode determined from the temperature dependent log J- V plots